\documentstyle[12pt]{article}

\def\p{\partial}
\begin{document}

\hfill

\vspace{4cm}

\centerline{\Large \bf  Symmetry Reductions of the Lax Pair}

\vspace{0.2cm}

\centerline{\Large \bf of the Four-Dimensional Euclidean}

\vspace{0.2cm}

\centerline{\Large \bf Self-Dual Yang-Mills Equations}

\vspace{0.5cm}

\centerline{M. LEGAR\'E}

\centerline{\footnotesize
Department of Mathematics, University of Alberta, Edmonton, T6G 2G1, Canada}

\vspace{3cm}

The reduction by symmetry of the linear system of the self-dual
Yang-Mills equations in four-dimensions under representatives of the
conjugacy classes of subgroups of the connected part to the identity of the
corresponding Euclidean group under itself is carried out. Only subgroups
leading to systems of differential equations nonequivalent to conditions of
zero curvature without parameter, or to systems of uncoupled first order
linear O.D.E.'s are considered. Lax pairs for a modified form of the Nahm's
equations as well as for systems of partial differential equations in two
and three dimensions are written out.

\vspace{1cm}

PACS(1995): 11.15.-q, 11.10.Lm

\newpage

{\bf I. INTRODUCTION}

Several systems of partial differential equations have been investigated in
the past via the method of symmetry reduction (see Refs.1--3, and references
therein). This includes the (coupled) Yang-Mills theories$^{4-9}$, and in
particular the self-dual Yang-Mills (abbreviated SDYM below) equations in
f{}lat spaces. In four dimensions, the latter equations are known to be
completely solvable through the twistor construction.$^{10,11}$ Their
reductions under symmetries, with often the addition of algebraic
constraints, have produced a large number of known integrable systems in
lower dimensions (for details, see Refs. 12 and 13), such as: the
Nahm$^{14-16}$, Boussinesq$^{17-19}$, (modified) Korteweg-de
Vries$^{17-19,20,21}$, (generalized) nonlinear Schr\"odinger$^{19-21}$,
N-wave$^{22}$ and Kadomtsev-Petviashvili$^{22}$ equations. Most of these
reductions have been accomplished using only translations, and their
hierarchies have been examined through the same reductions in Ref. 23.
Moreover, symmetry reductions using different invariant fields, or ansatze,
have been effected for higher dimensional versions of the SDYM
equations$^{24-26}$ as well as for some generalizations to self-dual
spaces$^{27}$.

The corresponding linear system, or Lax pair,$^{10,11,28,29}$ to the
SDYM equations has also been reduced with respect to translations as
well as other two- and three-dimensional Abelian subgroups of the
conformal group.$^{30-32}$ As expected, the compatibility of the
reduced Lax pair led to the SDYM equations reduced under the same
symmetry group. Let us mention that the six trancendents of
Painlev\'e were found in Ref. 32 as the result of reductions with
respect to three-dimensional Abelian subgroups, which have also been
derived through further reductions of reduced systems of the SDYM equations
(see Ref. 13). The symmetries involved consisted of translations,
rotations, and dilatations. In particular, the nontrivial reduction of the
Lax pair for the SDYM equations to the Lax pair for the Painlev\'e equation
$P_{VI}$ has been exhibited.

 In the following, the symmetry reduction of the Lax pair of the SDYM
equations with respect to any subgroup of the conformal group is
described.  In our discussion, we will restrict ourselves to the
Euclidean version of the SDYM system, where preliminary results have been
obtained $^{33}$. This work can also be performed on $R^4$ endowed with
the diagonal metric:  $(+1,+1,-1,-1)$ ($R^{(2,2)}$) $^{34,35}$. Let us
recall that the procedure of symmetry reduction has been applied to generate
new gauge invariant or supersymmetric systems from higher dimensional ones.
There is in general a residual gauge symmetry after reduction, but no
residual supersymmetry is ensured. Despite this result, supersymmetric
extensions of the SDYM equations in Euclidean space have been reduced by
subgroups of the Euclidean group (Ref. 36 and references therein) and
supersymmetric versions of known integrable systems have been produced.
Similarly, some superintegrable systems, such as the super-Korteweg-de
Vries and super-Toda field equations, have been recovered from the
supersymmetric SDYM equations in $R^{(2,2)}$ with the help of
differential constraints.$^{37}$ As a further motivation to this work,
let us mention that extended self-dual supersymmetric Yang-Mills theories
correspond to low energy limits of open or heterotic $N=2$ superstring
theories.$^{38,39}$

In order to introduce our notation, we recall in section II the
four-dimensional SDYM equations as well as their corresponding linear
system, or Lax pair, in Euclidean space ($E^4$). Then, the lift of
the action of the conformal group $SO(5,1)$, which leaves invariant
both the SDYM equations and its Lax pair is found by reference to the
twistor construction. section III reviews the invariance conditions for
the different elements involved in the linear system: i.e. the
Yang-Mills fields and the multiplet of scalar fields
(vector-functions), transforming under the fundamental representation
of the gauge group. A classification of the subalgebras of the real
Euclidean Lie algebra, $e(4) \sim so(4) \triangleright t^4$, of the
Euclidean group ($E(4)\sim O(4)\times \kern-12pt \supset  T^4$),
with respect to its
connected part to the identity ($E_o(4)$) is also indicated. We then
describe an algorithm of symmetry reduction of the Lax pair for the
SDYM equations and provide explicit examples in section IV. The
reduced Lax pairs obtained through reductions under representatives
of the conjugacy classes of subgroup of $E_o(4)$ giving rise to nontrivial
nonlinear differential systems of reduced equations are presented in
section V. We end this article with a summary of the results and some
comments regarding future directions of this work.

\vspace{1cm}

{\bf II. SELF-DUAL YANG-MILLS EQUATIONS AND LAX PAIR}

\vspace{0.3cm}

Let us write the SDYM equations in $E^4$ to set our notation:

$$ F= *F, \eqno (2.1)$$
where $F$ is a curvature 2-form pulled back to $E^4$ from the gauge bundle
$P(E^4,H)$, explicitly : $F=d\omega + \omega\wedge\omega$, with the
connection 1-form $\omega$ on $P$ taking values in the Lie algebra $\cal H$
of the gauge group $H$.

In terms of Cartesian coordinates $\{x^\mu\}$, they can be
expressed as:
$$F_{\mu\nu}=\frac{1}{2} \epsilon_{\mu\nu\kappa\sigma}F_{\kappa\sigma},
\eqno(2.2)$$
where $\mu,\nu,... = 1,...,4$, $\epsilon_{\mu\nu\kappa\sigma}$ stands
for the completely antisymmetric tensor in four dimensions with the
convention:  $\epsilon_{1234}=1$. The components of the field
strength ($F_{\mu\nu}$) are given by:
$$
F_{\mu\nu}=\p_\mu A_\nu-\p_\nu A_\mu + [A_\mu,A_\nu].
\eqno(2.3)
$$

The solutions to the SDYM equations on complexified Minkowski space
($M^C$), or self-dual connections on a vector bundle $N$ over
$M^C$, are related in a one-to-one manner to holomorphic vector
bundles $\tilde N$ over $CP^{3*}$, trivial over $CP^1$
submanifolds. The fibres ($C^n)$ of the bundle $\tilde N$ consist of
covariantly constant sections $\Psi$ of the bundle $N$
on anti-self-dual planes of
$M^C$ corresponding to a point of $CP^{3*}$, which are also called
$\beta$-planes. The condition of self-duality, or the self-dual
equations, of the connections on $M^C$ is in fact equivalent to the
condition of covariant constancy of  sections $\Psi$ with respect to the
(self-dual) connection on the anti-self-dual planes in $M^C$. The
latter condition can be interpreted as a Lax pair. By imposing a
suitable antiholomorphic involution on $CP^{3*}$, a fibration
$CP^{3*}\rightarrow S^4$ with
``real lines" $CP^1$ is induced, hence $CP^{3*}=CP^{3*}(S^4,CP^1)$. In the
same manner, the vector bundle over $S^4$ is pulled back to a
holomorphic vector bundle over $CP^{3*}$ if the self-duality
condition on $S^4$ is satisfied.$^{10, 11, 40}$
These constraints correspond to
a Lax pair and can be expressed as follows if we introduce a chart
$R^4$ of $S^4$ with coordinates $x^\mu$:
$$
[D_1+iD_2 -\lambda(D_3+iD_4)]\Psi (x,\lambda , \bar\lambda )=0,
\eqno (2.4a)
$$
$$
[D_3-iD_4 + \lambda(D_1-iD_2)]\Psi (x,\lambda ,\bar\lambda )=0,
\eqno (2.4b)
$$
$$
\p_{\bar \lambda} \Psi (x,\lambda , \bar\lambda )= 0,
\eqno (2.4c)
$$
where the covariant derivative: $D_\mu:= \p_\mu +A_\mu$ and $\lambda \in
CP^1$. The vector-function, or multiplet of scalar fields, $\Psi$ is a
holomorphic section of the vector bundle $\tilde N$ over $CP^{3\ast }$,
as expressed by (2.4c).

On the subset $R^4 \times R^2$ of $CP^{3*}$, labelled by the coordinates
$(x^\mu, y^i), i=1,2$, one finds that the vector parts of eqs (2.4):
$$\p_1+i\p_2-\lambda\p_3-i\lambda\p_4, \eqno(2.5a)$$
$$\lambda \p_1-i\lambda \p_2 +\p_3-i\p_4, \eqno (2.5b)$$
and $$\p_{\bar \lambda}=\frac{1}{2} \left (\frac{\p}{\p y^1}+ i \frac{\p}
{\p y^2} \right ), \eqno (2.5c)$$ define a basis of antiholomorphic vector
fields with respect to the complex structures $\cal J$ on $CP^{3*}$:
$$
{\cal J} = \{ J^{\nu }_{\mu }=-s_a \delta^{\nu\rho}\bar \eta^{a}_
{\rho\mu},\
\epsilon_{i}^{j}\},
\eqno (2.6)
$$
where the antisymmetric two-index tensor
$\epsilon_i^j$ ($i,j,... = 1,2$) is normalized to unity ($\epsilon_1^2=-1$),
$s_a $ ($a,b,... = 1,2,3$) are the Cartesian coordinates on $R^3$ such that
$s_as_a=1$, which parametrize the fibre $CP^1 \simeq S^2$, and a chart of
$S^2$ with variables $(y^1,y^2)$ where: $\lambda = y^1+iy^2$, has been
chosen via a stereographic projection. Explicitly, we have:
$$
s_1=\frac{2y^1}{1+|y|^2}=\frac{\lambda+\bar
\lambda}{1+\lambda\bar\lambda},
$$
$$
s_2=\frac{2y^2}{1+|y|^2}=\frac{i(\bar \lambda - \lambda)}{1+
\lambda\bar\lambda} ,
$$
$$
s_3=\frac{1-|y|^2}{1+|y|^2}=\frac{1-\lambda \bar\lambda}{1+\lambda
\bar\lambda},
\eqno (2.7)
$$
where $|y|^2=(y^1)^2+(y^2)^2$. The 't Hooft tensor
($\eta^a_{\mu \nu}$) and its dual ($\bar\eta^a_{\mu\nu}$) are given by (see
Ref. 41 for identities):
$$\eta^a_{bc}=\bar\eta^a_{bc}=\epsilon_{abc},
\eqno(2.8a)
$$
where $a,b,c=1,2,3$ and $\epsilon_{abc}$ is the three-dimensional
antisymmetric tensor ($\epsilon_{123}=1$),
$$\bar\eta^a_{b4}=-\eta^a_{b4}=-\delta^a_b, \eqno (2.8b)$$
$$\eta^a_{\mu\nu}=-\eta^a_{\nu\mu}, \eqno (2.8c)$$
$$\bar\eta^a_{\mu\nu}=-\bar\eta^a_{\nu\mu}. \eqno (2.8d)$$

The SDYM equations (2.2) and their linear system (2.4) are invariant
under the gauge transformations:
$$
A_\mu^\prime= h^{-1}A_\mu h +h^{-1}\p_\mu h,
\eqno(2.9a)
$$
and
$$ \Psi^\prime= h^{-1}\Psi ,
\eqno (2.9b)
$$
where $h \in H$ is a function of $x \in S^4$. These
equations are also preserved by the global action of the conformal
group $SO(5,1)$. In order to preserve the holomorphic structure of the
bundle $\tilde N\rightarrow CP^{3\ast }$, the action of $SO(5,1)$
is lifted to $CP^{3\ast }$ in a holomorphic fashion by requiring the
complex structure (2.6) to be invariant with respect to a lifted action of
the conformal group. Locally, the lifted vector fields
($\tilde X$)  will obey to the Lie algebra $so(5,1)$ of $SO(5,1)$,
and will correspond to infinitesimal automorphisms of the complex
structure (2.6), i.e. $^{42,43}$ :
$$
{\cal L}_{\tilde X} {\cal J} = 0,
\eqno(2.10)
$$
$\forall X \in so(5,1)$, where ${\cal L}_{\tilde
X}$ denotes the Lie derivative with respect to $\tilde X$.

A specific representation of $so(5,1)$ can be realized in terms of vector
fields ($\hat X$) on $E^4$
$$
\hat X_a=-\frac{1}{2}\delta_{ab}\eta^b_{\mu\nu} x_\mu\p_\nu,
\quad
\hat Y_a=-\frac{1}{2}\delta_{ab}\bar\eta^b_{\mu\nu}x_\mu\p_\nu,
\quad
\hat P_\mu=\p_\mu,
$$
$$
\hat K_\mu =\frac{1}{2}x_\sigma
x_\sigma \p_\mu-x_\mu \hat D, \quad \hat D = x_\sigma \p_\sigma,
\eqno(2.11)
$$
where $\{\hat X_a, a=1,2,3\}$ and $\{\hat Y_a, a=1,2,3\}$ are two commuting
$so(3)$ Lie algebras of $so(4)$, $\hat K_\mu$ denotes the generators of the
special conformal transformations, and $\hat D$ generates the dilatations.

One verifies that the lifted vector fields on $R^4 \times CP^{1}=
CP^{3*}\backslash CP^1$ can be expressed as:
$$
\tilde X_a=\hat X_a,\quad \tilde Y_a= \hat Y_a - Z_a, \quad
\tilde P_\mu=\hat P_\mu,
$$
$$
\tilde K_\mu=\hat K_\mu + \bar\eta^a_{\sigma\mu}x_\sigma Z_a, \quad \tilde
D=\hat D, \eqno(2.12)
$$
with the generators $\{Z_a,a=1,2,3\}$ forming the
$SO(3)$ rotations on $S^2$, or vector fields $\in T(CP^{1})$ such that:
$$Z_a=\epsilon_{abc}s_b\p_c, \eqno(2.13)$$
which in terms of the parameter $\lambda$ become:
$$Z_1=\frac{i}{2}[(\lambda^2-1) \p_\lambda + (1-{\bar\lambda}^2) \p_{\bar
\lambda}],
$$
$$Z_2=\frac{1}{2}[(1+\lambda^2)\p_\lambda + (1+{\bar\lambda}^2)
\p_{\bar\lambda}],$$
$$Z_3=i(\lambda \p_\lambda - \bar\lambda \p_{\bar\lambda}).
\eqno(2.14)
$$

Let us restrict ourselves to the (real) Euclidean Lie algebra $e(4) \sim
so(4) \triangleright t^4$, which can be realized as an embedding in
$so(5,1)$. We introduce the matrix $I_{5,1}$, defined as:
$$I_{5,1}= \left [\matrix {&&&\vdots\quad\quad&\cr
&{\bf1}_{4}&&\vdots\quad\,\,\,0 \cr
&&&\vdots\quad\quad&\cr
\ldots&\ldots&\ldots&\ldots&\ldots\cr &&&\vdots\,\,-1&0
\cr &0&&\vdots\quad\quad\cr  &&&\vdots\quad\,\,\,0&1
\cr} \right ].
\eqno(2.15)
$$

Then $so(5,1)$ consists of the set of elements $S \in gl(6,R)$ satisfying
the relation :
$$ S^T I_{5,1} + I_{5,1} S = 0. \eqno(2.16)$$
Among those elements, the algebra $e(4)$ is determined by the subset
composed of: $$\left [\matrix {&&\vdots\quad\quad\cr
&{\bf Y}&\vdots\quad\,\,\,0\cr
&&\vdots\quad\quad\cr
\ldots&\dots&\ldots\,\,\ldots\cr
&&\vdots\quad\quad\cr &0&\quad\vdots\quad{\bf
0}_2\quad\cr        &&\vdots\quad\quad\cr} \right
] \quad, \quad\left [\matrix
{&&&\quad\quad\,\,\,\vdots\,\,\alpha_1&\alpha_1\cr
&&&\quad\quad\,\,\,\vdots\,\,\alpha_2&\alpha_2\cr
&&{\bf0}_{4}&\quad\quad\,\,\,\vdots\,\,\alpha_3&\alpha_3\cr
&&&\quad\quad\,\,\,\vdots\,\,\alpha_4&\alpha_4&\cr
\ldots&\ldots&\ldots&\ldots&\ldots\cr
\alpha_1&\alpha_2&\alpha_3&\alpha_4\,\,\,\,\,\vdots\,\,0&0\cr
-\alpha_1&-\alpha_2&-\alpha_3& -\alpha_4\vdots\,\,0&0\cr}
\right ],
\eqno(2.17)
$$
where $Y^T=-Y$, each element belonging to $so(4) \sim so(3)\oplus so(3)$,
and the translations ($P_\mu$) along the $x^\mu$-axis, parametrized by
$\alpha_\mu$.

The linear action of the Euclidean group on $R^6$ provided with the
Cartesian coordinates $(\eta^1,...,\eta^6)$ determines the standard action
of $E_o(4)$ on $E^4$ through the formula:
$$x^\mu=\frac {\eta^\mu}{\eta^5+\eta^6}. \eqno(2.18)$$

We have elected the following basis \footnote{The notation
$A_1,A_2,A_3,B_1,B_2,$ and $B_3$ was previously used in reference 44.} of
$so(4) \subset gl(6,R)$:
$$
A_1=-X_3=\frac{1}{2}(M_{12}+M_{34}),\;
A_2=-X_1=\frac{1}{2}(M_{23}+M_{14}),\;
A_3=X_2=\frac{1}{2}(M_{13}-M_{24}),
\eqno(2.19)
$$ $$
B_1= - Y_3=\frac{1}{2}(M_{12}-M_{34}), \;
B_2= - Y_1=\frac{1}{2}(M_{23}-M_{14}), \;
B_3=Y_2=\frac{1}{2}(M_{13}+M_{24}),
\eqno(2.20)
$$ $$
[M_{\alpha\beta}]_{\mu\nu}=\delta_{\mu\alpha}\delta_{\nu\beta} -
\delta_{\mu\beta}\delta_{\nu\alpha},\quad
[M_{\alpha\beta}]_{56}=0,
\eqno(2.21)
$$
where $\alpha, \beta, \mu, \nu
=1,..,4$. Let us note that $M_{\alpha\beta}$ generate rotations in the
$(x^{\alpha}, x^{\beta})$-plane.

Since the SDYM equations and their Lax pair are left unchanged by the
action of $E_o(4)$, reductions with respect to subgroups conjugated
under $E_o(4)$ will produced equivalent reduced systems (see
Refs.1--3). We can therefore limit ourselves to reductions by a
subgroup representative of each conjugacy class of subgroups of
$E_o(4)$.$^{44}\ $  Such a classification has been carried out for
different subalgebras of interest in physics (see Refs. 3, 44--47 and
references therein) and a ``normalized" list of representatives of
the conjugacy classes of subalgebras of $e(4)$ under $E_o(4)$ has
been obtained in Ref. 44.

\vspace{1cm}

{\bf III. INVARIANCE CONDITIONS}

\vspace{0.3cm}

The linear system (2.4) of the SDYM equations involves Yang-Mills fields
($A_\mu$) and multiplets ($\Psi$) of scalar fields transforming under the
fundamental representation of the gauge group $H$.

The Yang-Mills fields can be interpreted as pullbacks to
the base manifold of connection 1-forms on $P(E^4, H)$, and their invariance
has been studied in many papers. For instance, one may consult Refs.5--8.

All the isotropy subgroups of the representatives of the conjugacy
classes with orbits of dimension one, two, or three correspond either to
the identity or to a compact Lie group: $SO(2)$ or $SO(3)$, and the
approach presented in Refs. 5,6 and 8 can then be used to determine the most
general and globally invariant gauge fields in $E^4$. However, we are only
interested in local expressions for the symmetric fields, and we will
impose the infinitesimal form of these conditions.

Let us suppose that the symmetry group $G$ acts (effectively) on each orbit
$G/G_o$ with cross-section $V$, where $G_o$ is identified as the isotropy
subgroup of $G$ at each point of $V$.

In finite form, the invariance conditions are given by $^{5,6,8}$ :

 $$f^*_g\omega=\rho^{-1}(g,x)\,\omega\,\rho(g,x) + \rho^{-1}
(g,x)d\,\rho(g,x),
\eqno(3.1)$$
where $\omega^\sigma=A_\mu\,dx^\mu$.

Infinitesimally, we have that :
$${\cal L}_{\tilde X}\omega^\sigma=DW:=dW+[\omega^\sigma,W], \eqno(3.2)$$
$\forall g \in G$, or  $\forall X \in {\cal G}$ (the Lie algebra of $G$),
where the gauge transformation $\rho : G \times E^4 \rightarrow H$,
specifies the lift of the group action  ($f_g$) to the gauge bundle over
$G/G_o \times V$ (cf  Refs 5, 6 and 8). The latter equation can be
interpreted as the vanishing of the Lie derivative of the Yang-Mills fields
with respect  to a vector field
$\tilde X$ induced by an element $X$ of the symmetry algebra ${\cal G}$  up
to an infinitesimal gauge transformation, where $W: {\cal G} \times E^4
\rightarrow {\cal H}$(the Lie algebra of $H$). Let us point out that the
map $W$ is defined as:
$$W=\frac{d}{dt} \rho(g=e^{tX},x)|_{t=0}, \eqno(3.3)$$
Its vanishing leads to a strict invariance condition, i.e. an invariance of
the field without the help of any gauge transformation.

As for the multiplet of scalar fields, their finite and infinitesimal
invariance conditions can be respectively read as:
$$\tilde f^*_g\Psi = \rho^{-1}(g,x)\,\Psi, \eqno(3.4)$$
and
$${\cal L}_{\tilde X^\ast}\Psi = -W\Psi, \eqno(3.5)$$
$\forall X \in {\cal G}$, where $\tilde f^*_g$ and $\tilde X^\ast$ are
respectively
the lifts of the action and of the vector field associated to $X$.

For simplicity, all the cases below involve only an invariance without
gauge transformation. Other reduced systems might be derived by
substitution in the SDYM equations of invariant fields solutions to (3.1),
(3.4) or (3.2), (3.5) with non-vanishing $\rho$ and $W$ functions (cf Ref.
33).

\vspace{1cm}

{\bf IV. REDUCTION BY SYMMETRY OF THE LAX PAIR}

\vspace{0.3cm}

Let us consider a symmetry group $G$, subgroup of the invariance
group $SO(5,1)$. The procedure of symmetry reduction consists essentially
in substituting $G$-invariant $A_\mu$ and $\Psi$ on $S^4 \times CP^1$ in
the set of differential equations, rewritten in terms of the orbit and
invariant coordinates of the $G$-action. Once a basis of a $n$-dimensional
representative
${\cal G}$ of a conjugacy class of the subalgebras of $e(4)$ is chosen:
$\{X_i, i=1,...,n\}$, we first determine its induced vector fields on $R^4
\times CP^1 \subset S^4 \times CP^1$: $\{\tilde X_i, i=1,...,n\}$,
then select orbit variable(s): $\{\xi_m, m=1,...,\leq n\}$, normally
group parameters, and determine the invariant coordinates: $\{\chi_A\}$,
with the formula:  $${\cal L}_{\tilde X_i}\chi_A=0, \eqno(4.1)$$
$\forall i=1,...,n.$ Among the invariant variables $\chi_A$, we can
identify a (new) spectral parameter, denoted $\zeta$, which obeys to
${\cal L}_{\hat X_i}\zeta\neq0$, for some $i=1,...,n$.

After insertion in the Lax pair (2.4) of the coordinates $(\xi_m,
\chi_A)$ and the $G$-invariant fields $A_\mu$ and $\Psi$, obtained via the
method presented in the previous section, we are left using the
holomorphicity condition (2.4c) and by elimination of multiplying factors
functions of the orbits coordinates, with two reduced equations depending
solely on invariant variables.

We can rewrite these reduced equations by dividing their differential and
potential parts as:
$$\nabla_X\Psi = (X+A_X)\Psi=0, \eqno(4.2a)$$
and
$$\nabla_Y\Psi = (Y+A_Y)\Psi=0, \eqno(4.2b)$$
where $X$ and $Y$ represent the respective vector components of
the reduced equations of the Lax pair (2.4).

It can be verified that the compatibility of the system (4.2) coincide with
the SDYM equations reduced under the same subgroup $G$:
$$[\nabla_X,\nabla_Y] - \nabla_{[X,Y]}=0. \eqno(4.3)$$

Let us add for a simple treatment that the equation (2.4c) can be
interpreted as an invariance of $A_\mu$ (trivially) and $\Psi$ under the
translations ($P_{\bar\lambda}$) along the complex coordinate
$\bar\lambda$ on $R^2 \subset CP^1$.

However, the compatibility (4.3) of (4.2) does not necessarily tally with
the reduced SDYM equations if the residual vector components span only a
one dimensional vector space. Still, the equations (4.2) respect the
equality:
$$[\nabla_X,\nabla_Y]\Psi=f_X\nabla_X\Psi + f_Y\nabla_Y\Psi, \eqno(4.4)$$
where $f_X$ and $f_Y$ are functions of the invariant coordinate(s), which
give rise to the residual SDYM equations when appropriately chosen. A
reduced Lax pair, producing the correct SDYM equations after reduction with
respect to the same symmetry group is obtained if certain multiplying
factors are adjoined to each of the operators $\nabla_X$ and $\nabla_Y$.

In fact, we have:
$$[h_X\nabla_X,h_Y\nabla_Y]=0, \eqno(4.5)$$
with $h_X$ and $h_Y$, functions of the invariant coordinates.

{}From the commutator (4.5), we deduce that:
$$[X,Y] +\frac{1}{h_Y}(Xh_Y)Y-\frac{1}{h_X}(Yh_X)X=0, \eqno(4.6a)$$
and
$$XA_Y-YA_X+[A_X,A_Y] + \frac{1}{h_Y}(Xh_Y)A_Y-\frac{1}{h_X}(Yh_X)A_X=0.
\eqno(4.6b)$$
The functions $h_X$ and $h_Y$ are determined by solving (4.6) with the
requirement that (4.6b) corresponds to the reduced SDYM equations. Let us
indicate that holonomic vector components to any reduced Lax pair can be
found by introducing the above-mentioned factors : $h_X$ and $h_Y$. For
two- and three-dimensional vector fields, holonomic components can be
determined by solving uniquely (4.6a), then (4.6b) will automatically
coincide with the reduced SDYM equations.

We end this section by presenting two examples which illustrates the above
method :

\vspace{.5cm}

(1) $\{Y_3, P_3, P_4\}$:

The lifted vector fields have the form:  $$\tilde
Y_3=-\frac{1}{2}(x^1\p_2-x^2\p_1-x^3\p_4+x^4\p_3)-i(\lambda\p_\lambda
- \bar\lambda\p_{\bar\lambda}) = -\frac{1}{2} \p_\varphi, \eqno(4.7a)$$
with $$\tilde P_3=\p_3, \eqno(4.7b)$$ $$\tilde P_4=\p_4.
\eqno(4.7c)$$

Since the lift of $Y_3$ to $CP^{3*}$ is nontrivial, we expect a new
spectral parameter among the invariant variables. The orbit
coordinates are:  $$\varphi = -\arctan (\frac{x^2}{x^1}) +
\frac{i}{4}\ln(\frac{\lambda} {\bar\lambda}), x^3, x^4 \eqno(4.8)$$
and the invariant coordinates can be chosen as: $$
r=\sqrt{(x^1)^2+(x^2)^2}, \Lambda = \lambda \bar\lambda$$ and $$\eta
= -\arctan (\frac{x^2}{x^1}) - \frac{i}{4}\ln(\frac{\lambda}
{\bar\lambda}) \eqno(4.9)$$

In terms of these variables on the stratum, the symmetric Yang-Mills fields
have the form:
$$(A_1, A_2, A_3, A_4)^T=e^{-2\,\theta Y_3}(u_1, u_2, u_3, u_4)^T,
\eqno(4.10)$$
where $\theta = \frac{\eta+\varphi}{2}$, $u_\mu =u_\mu (r),
\mu=1,...,4$ and $\Psi=\psi(r, \Lambda, \eta)$.

Inserting (4.8), (4.9), (4.10) and $\Psi$ in the linear system (2.4),
we find:
$$[\p_r-\frac{i}{r}\p_\eta + u_1+iu_2 - e^{i2\theta} \lambda
(u_3+iu_4)]\psi = 0,
\eqno(4.11a)
$$ $$
[e^{i2\theta} \lambda (\p_r + \frac{i}{r} \p_\eta + u_1-iu_2)
+ u_3-iu_4]\psi = 0.
\eqno(4.11b)
$$

The condition (2.4c): $\p_{\bar\lambda}\Psi = 0$ or $(\frac{\p}{\p
\sqrt{\Lambda}} + \frac{i}{2 \sqrt{\Lambda}}\frac{\p}{\p \eta})\psi =
0$, restricts us to the invariant $\zeta= \sqrt{\Lambda} e^{i2\eta}$.
In terms of the new spectral parameter $\zeta$, the equations (4.11)
become the reduced Lax pair:
$$\nabla_X\Psi=[\p_r+\frac{2\zeta}{r}\p_\zeta +
u_1+iu_2-\zeta(u_3+iu_4)]\psi=0, \eqno(4.12a)$$
$$\nabla_Y\Psi=[\zeta(\p_r-\frac{2\zeta}{r}\p_\zeta)+
\zeta(u_1-iu_2)+(u_3-iu_4)]\psi=0,
\eqno(4.12b)
$$
A Lax pair expressed in terms of holonomic vector fields is derived
from (4.12) if $h_X\propto r$ and $h_Y\propto \frac {r}{\zeta}$. The
SDYM equations reduced under the same subgroup arise as the
compatibility of the linear system (4.12):$^{44}$
$$\dot u_2
+\frac{u_2}{r} +[u_1,u_2]-[u_3,u_4] =0,$$ $$\dot u_3 -\frac{u_3}{r}
+[u_1,u_3]+[u_2,u_4] =0, \eqno(4.13)$$ $$\dot u_4 -\frac{u_4}{r}
+[u_1,u_4]+[u_3,u_2] =0,
$$
where a dotted variable indicates a
differentiation with respect to its argument. With the change of
variables: $\xi=\ln (r), w_2=r\,u_2, w_3=r^{-1}u_3, w_4=r^{-1}u_4$,
and the gauge condition $u_1=0$, the integrability of (4.12) leads to
a modified form of the Nahm's equations.$^{44}$

To simplify computations in Example 2, the vector $P_{\bar\lambda} =
\p_{\bar\lambda}$ of the holomorphicity equation (2.4c) is included in the
symmetry algebra of lifted vector fields.

\vspace{.5cm}

(2) $\{X_3,Y_3\}$:

Using the holomorphicity condition or invariance along $\bar\lambda$:
$\p_{\bar\lambda}\Psi=0$:
$$\tilde X_3= -\frac{1}{2}(x^1\p_2-x^2\p_1+x^3\p_4-x^4\p_3) =
-\frac{1}{2}\p_\chi, \eqno(4.14a)$$
$$\tilde Y_3= -\frac{1}{2}(x^1\p_2-x^2\p_1-x^3\p_4+x^4\p_3) -
i(\lambda\p_{\lambda} - \bar\lambda\p_{\bar\lambda}) = \frac{1}{2}\p_\phi,
\eqno(4.14b)$$
where we have elected the orbit coordinates: $\chi$, $\phi$, and
$\bar\lambda$ and the invariant variables: $r, R$, and $\zeta$,
which correspond to: $$x^1=r \cos(\frac{\chi+\phi}{2}), x^2=-r
\sin(\frac{\chi+\phi}{2}),$$ $$x^3=R \cos(\frac{\chi-\phi}{2}), x^4=-R
\sin(\frac{\chi-\phi}{2}), \lambda = e^{2i\phi}\zeta, \eqno(4.15)$$
Here $\zeta$ stands for the new spectral parameter.

The invariant Yang-Mills field obeying to (3.1) is given by:
$$(A_1, A_2, A_3, A_4)^T = e^{-\chi X_3} e^{-\phi Y_3}  (u_1, u_2, u_3,
u_4)^T,
\eqno(4.16)$$
where  $u_\mu = u_\mu (r, R)$ and $\Psi=\psi(r, R, \zeta)$.

Substitution of (4.15), (4.16), and $\Psi$ into the Lax pair (2.4) implies
that:
$$\nabla_X\Psi= \left [\p_r - \zeta\p_R + \left ( \frac {\zeta}{r} +
\frac{\zeta^2}{R} \right )\p_\zeta + u_1+iu_2-\zeta(u_3+iu_4) \right ]
\psi=0,
\eqno(4.17a)$$
$$\nabla_Y\Psi= \left [\p_R + \zeta\p_r + \left ( \frac {\zeta}{R} -
\frac{\zeta^2}{r} \right )\p_\zeta + u_3-iu_4+\zeta(u_1-iu_2) \right ]
\psi=0.
\eqno(4.17b)$$
The reduced SDYM equations are recovered through the compatibility of the
system (4.17) and have the form:
$$\p_r u_3 - \p_R u_1 + [u_1,u_3] + [u_2,u_4] = 0, \eqno(4.18a)$$
$$\p_r u_4 + \p_R u_2 - [u_2,u_3] + [u_1,u_4] = 0, \eqno(4.18b)$$
$$\p_R u_4 - \p_r u_2 + \frac{u_4}{R}-\frac{u_2}{r} - [u_1,u_2] +
[u_3,u_4]=0. \eqno(4.18c)$$
A Lax pair with holonomic vectors follows if we put $h_X\propto r$ and $h_Y
\propto R$.

\vspace{1cm}

{\bf V. REDUCED LAX AND SDYM EQUATIONS}

\vspace{0.3cm}

In this section, we present the resulting symmetry reductions with
respect to the representatives of the classes of subalgebras of
$e(4)$ (see Table 2 of Ref. 44) giving rise to differential
systems which are not equivalent to a zero curvature (without parameter)
condition on the residual potentials, since the latter are then gauge
equivalent to vanishing solutions (see for instance the subalgebra 4d),  or
to systems of uncoupled first order linear O.D.E.'s after a gauge choice
(see for instance the representative 6b). One finds the list of
representatives of these subalgebras in Table 1 of this article.  For
comparison, the Ref. 44 includes  all the reduced SDYM equations under the
representatives of $e(4)$. The labels attached to the representatives refer
to the numbering adopted in Ref. 44, the generators or elements forming a
basis are also specified within curly brackets. We then provide the orbit
and invariant variables, as well as the invariant Yang-Mills fields
determined according to section III.  The reduced Lax pairs and their
compatibility condition, the reduced SDYM equations, follow. In order to
simplify the computations, we have considered the vector
$P_{\bar\lambda} = \p_{\bar\lambda}$ of the holomorphicity equation
(2.4c) as part of the symmetry algebra of lifted vector fields. This
extension of the representative gives rise to the same reduced equations
since the residual symmetry of the reduced equations includes $\{
\p_\lambda,
\p_{\bar\lambda} \}$ . We have parametrized the orbits of the translations
generated under $P_{\bar\lambda}$ with the coordinate $\bar\lambda$. In the
following, the reduced equations are first written without any
special choice of gauge. In some cases, a relation to already known
integrable systems is indicated.
\vspace{.5cm}

{\bf 1.} 1a $\{P_4\}$
\vspace{.5cm}

Orbit coordinates: $x^4, \bar\lambda$

Invariant coordinates: $x^1, x^2, x^3, \lambda$

Invariant $A_\mu$ and $\Psi$:
$$A_\mu=u_\mu(x^1, x^2, x^3), \Psi=\psi(x^1, x^2, x^3, \lambda)
\eqno(5.1)$$

Reduced Lax pair:
$$[\p_1+i\p_2+u_1+iu_2-\lambda(\p_3+u_3+iu_4)]\psi=0$$
$$[\p_3+u_3-iu_4+\lambda(\p_1-i\p_2+u_1-iu_2)]\psi=0 \eqno(5.2)$$

Reduced SDYM equations:
$$\p_1u_2-\p_1u_2-\p_3u_4+[u_1,u_2]-[u_3,u_4] = 0$$
$$\p_2u_3-\p_3u_2-\p_1u_4+[u_2,u_3]-[u_1,u_4] = 0 \eqno(5.3)$$
$$\p_1u_3-\p_3u_1+\p_2u_4+[u_1,u_3]+[u_2,u_4] = 0,$$

which correspond to the Bogomolny equations $^{48,49}$ with  $u_4 = \phi$.

\vspace{.5cm}

{\bf 2.} 1b $\{P_3,P_4\}$
\vspace{.5cm}

Orbit coordinates: $x^3, x^4, \bar\lambda$

Invariant coordinates: $x^1, x^2, \lambda$

Invariant $A_\mu$ and $\Psi$:
$$A_\mu=u_\mu(x^1, x^2), \Psi=\psi(x^1, x^2, \lambda) \eqno(5.4)$$

Reduced Lax pair:
$$[\p_1+i\p_2-\lambda(u_3+iu_4)+u_1+iu_2]\psi=0$$
$$[\lambda(\p_1-i\p_2+u_1-iu_2)+u_3-iu_4]\psi=0 \eqno(5.5)$$

Reduced SDYM equations:
$$\p_1u_2-\p_1u_2+[u_1,u_2]-[u_3,u_4] = 0$$
$$\p_2u_3-\p_1u_4+[u_2,u_3]-[u_1,u_4] = 0 \eqno(5.6)$$
$$\p_1u_3+\p_2u_4+[u_1,u_3]+[u_2,u_4] = 0$$

A number of algebraic reductions have been performed with the help of gauge
choices starting from eq.(5.6). For instance, one can find a reduction  to
the Toda lattice equations, to the chiral field equations (if null
variables are used), as well as the elliptic sine-Gordon equation $^{50}$.

\vspace{.5cm}
{\bf 3.} 1c $\{P_1,P_2,P_3\}$
\vspace{.5cm}

Orbit coordinates: $x^1, x^2, x^3, \bar\lambda$

Invariant coordinates: $x^4, \lambda$

Invariant $A_\mu$ and $\Psi$:
$$A_\mu=u_\mu(x^4), \Psi=\psi(x^4, \lambda) \eqno(5.7)$$

Reduced Lax pair:
$$[\lambda(i\p_4+u_3+iu_4)-u_1-iu_2]\psi=0$$
$$[i\p_4-\lambda(u_1-iu_2)-u_3+iu_4]\psi=0 \eqno(5.8)$$

Reduced SDYM equations:
$$\p_4u_1+[u_2,u_3]-[u_1,u_4] = 0$$
$$\p_4u_2-[u_1,u_3]-[u_2,u_4] = 0 \eqno(5.9)$$
$$\p_4u_3+[u_1,u_2]-[u_3,u_4] = 0$$

The Nahm equations$^{51-53}$ are recovered with the gauge
choice $u_4=0$ and $u_a\rightarrow  -u_a$.
\vspace{.5cm}

{\bf 4.} 2a,3a,4a,5a $\{\alpha X_3+\beta Y_3 | \alpha, \beta \in R\}$
\vspace{.5cm}

Orbit coordinates:
$\xi=-(\alpha+\beta)\arctan(\frac{x^2}{x^1})-(\alpha-\beta)
\arctan(\frac{x^4}{x^3}), \bar\lambda$

Invariant coordinates:
$\chi=(\alpha-\beta)\arctan(\frac{x^2}{x^1})-(\alpha+\beta)
\arctan(\frac{x^4}{x^3}), r=\sqrt{(x^1)^2 +(x^2)^2}$, $R=\sqrt{(x^3)^2
+(x^4)^2}$, $\zeta = e^{(i\gamma(-(\alpha+\beta)
\arctan(\frac{x^2}{x^1})-(\alpha-\beta) \arctan(\frac{x^4}{x^3})))} \lambda
= e^{i\gamma\xi} \lambda$, where $\gamma = \frac{\beta}{\alpha^2+\beta^2}$

Invariant $A_\mu$ and $\Psi$:
$$A_1=u_1 \cos\theta +  u_2 \sin\theta,\quad
A_2=-u_1 \sin\theta + u_2\cos\theta ,
$$
$$A_3=u_3 \cos\phi + u_4\sin\phi ,\quad
A_4=-u_3 \sin\phi + u_4\cos\phi ,
\eqno(5.10)$$ with
$\phi=\frac{(\alpha+\beta)\chi+(\alpha-\beta)\xi}{2(\alpha^2+\beta^2)},
\theta=\frac{-(\alpha-\beta)\chi+(\alpha+\beta)\xi}{2(\alpha^2+\beta^2)},
u_\mu=u_\mu(r,R,\chi)$ and $\Psi=\psi(r,R,\chi,\zeta)$
\vspace{.5cm}

Reduced Lax pair:
$$[ \p_r-\zeta e^{-i\Gamma \chi}\p_R +i
\left (\frac{\alpha-\beta}{r}+\frac{(\alpha+\beta)}{R} \zeta e^{-i\Gamma
\chi} \right ) \p_\chi \hspace{6cm}$$
$$+\left ( \frac{\gamma(\alpha+\beta)}{r}-\frac{\gamma (\alpha-\beta)}{R}
\zeta e^{-i\Gamma \chi} \right )\zeta\p_\zeta +u_1+iu_2 - \zeta e^{-i\Gamma
\chi}(u_3+iu_4) ] \psi = 0 \eqno(5.11a)$$
$$[ \p_R+\zeta e^{-i\Gamma \chi}\p_r +i
\left (\frac{\alpha+\beta}{R}-\frac{(\alpha-\beta)}{r} \zeta e^{-i\Gamma
\chi} \right ) \p_\chi \hspace{6cm}$$
$$-\left ( \frac{\gamma(\alpha-\beta)}{R}+\frac{\gamma (\alpha+\beta)}{r}
\zeta e^{-i\Gamma \chi} \right ) \zeta\p_\zeta +u_3-iu_4 + \zeta e^{-i
\Gamma
\chi}(u_1-iu_2) ] \psi = 0 \eqno(5.11b)$$
where $\gamma=\frac{\beta}{\alpha^2+\beta^2}$ and
$\Gamma=\frac{\alpha}{\alpha^2+\beta^2}$.

Let us note that holonomic vector parts can be obtained if (5.11a) and
(5.11b) are respectively multiplied by $r$ and $R$.

Reduced SDYM equations:
$$\p_Ru_4-\p_ru_2+\frac{u_4}{R}-\frac{u_2}{r}+
\frac{(\alpha+\beta)}{R}\p_\chi u_3+\frac{(\alpha-\beta)}{r}\p_\chi u_1
+[u_2,u_1] + [u_3,u_4] = 0$$
$$\p_ru_3-\p_Ru_1+\frac{(\alpha-\beta)}{r} \p_\chi
u_4+\frac{(\alpha+\beta)}{R} \p_\chi u_2 + [u_1,u_3]+[u_2,u_4] = 0
\eqno(5.12)$$
$$\p_ru_4+\p_Ru_2-\frac{(\alpha-\beta)}{r} \p_\chi
u_3+\frac{(\alpha+\beta)}{R} \p_\chi u_1+ [u_2,u_3] - [u_1,u_4] = 0
$$

\vspace{.5cm}

{\bf 5.} 2b,3b,4c,5b $\{\alpha X_3 + \beta Y_3, P_3, P_4 | \alpha, \beta
\in R\}$
\vspace{.5cm}

Orbit coordinates: $\theta=-\arctan(\frac{x^2}{x^1})$, $x^3, x^4,
\bar\lambda$

Invariant coordinates: $r,\zeta=e^{i\gamma\theta} \lambda$, where $\gamma=
\frac{2\beta}{\alpha+\beta}$

Invariant $A_\mu$ and $\Psi$:
$$A_1=u_1 \cos\theta + u_2\sin\theta ,\quad
A_2=-u_1 \sin\theta + u_2\cos\theta ,$$
$$A_3=u_3 \cos(\frac{\alpha-\beta}{\alpha+\beta}\theta) +
 u_4\sin(\frac{\alpha-\beta}{\alpha+\beta}\theta),\
A_4=- u_3\sin(\frac{\alpha-\beta}{\alpha+\beta}\theta) +
 u_4 \cos(\frac{\alpha-\beta}{\alpha+\beta}\theta)
\eqno(5.13)
$$
where $u_\mu = u_\mu (r)$ and $\Psi=\psi(r,\zeta)$.

Reduced Lax pair:
$$[\p_r+\frac{\gamma}{r}\zeta \p_\zeta + u_1+iu_2 - \zeta(u_3+iu_4)] \psi =
0$$
$$[\zeta \p_r - \frac{\gamma}{r} \zeta^2 \p_\zeta +\zeta(u_1-iu_2) +
u_3-iu_4]
\psi = 0 \eqno(5.14)$$

Holonomic vector components can be obtained if the two above equations are
multiplied by $r$.

Reduced SDYM equations:
$$\frac{du_2}{dr} + \frac{u_2}{r} + [u_1,u_2] - [u_3,u_4] = 0$$
$$\frac{du_3}{dr} +\frac{1-\gamma}{r} u_3 + [u_1,u_3] + [u_2,u_4] = 0
\eqno(5.15)$$
$$\frac{du_4}{dr} +\frac{1-\gamma}{r} u_4 + [u_1,u_4] + [u_3,u_2] =0$$

With a suitable change of variables and gauge choice, the above SDYM and
corresponding Lax equations can be algebraically reduced respectively to
the equations of the Toda lattice with damping and its linear system
$^{33}$.
\vspace{.5cm}

{\bf 6.} 4b  $\{X_3 + Y_3,P_3\}$
\vspace{.5cm}

Orbit coordinates: $\theta=-\arctan(\frac{x^2}{x^1}), x^3, \bar\lambda$

Invariant coordinates: $r=\sqrt{(x^1)^2+(x^2)^2}, x^4, \zeta=
e^{i\theta}\lambda$

Invariant $A_\mu$ and $\Psi$:
$$A_1=u_1 \cos\theta + u_2\sin\theta ,\
A_2=-u_1 \sin\theta + u_2\cos\theta , \
A_3 = u_3,\  A_4 = u_4
\eqno(5.16)
$$
where $u_\mu = u_\mu (r, x^4)$ and $\Psi=\psi(r, x^4, \zeta)$.

Reduced Lax pair:
$$[\p_r+\frac{1}{r}\zeta \p_\zeta -i\zeta\p_4 + u_1+iu_2 -
\zeta(u_3+iu_4)] \psi = 0$$
$$[\zeta \p_r - \frac{1}{r} \zeta^2 \p_\zeta - i\p_4 +\zeta(u_1-iu_2) +
u_3-iu_4] \psi = 0 \eqno(5.17)$$

One can find holonomic vector components if the two above equations are
respectively  multiplied by $r$ and $\frac{r}{\zeta}$.

Reduced SDYM equations:
$$\p_r u_2 + \p_4 u_3 + \frac{u_2}{r} + [u_1,u_2] - [u_3,u_4] = 0$$
$$\p_r u_3 - \p_4 u_2 + [u_1,u_3] + [u_2,u_4] = 0 \eqno(5.18)$$
$$\p_r u_4 - \p_4 u_1 + [u_1,u_4] + [u_3,u_2] = 0$$

\vspace{.5cm}

{\bf 7.} 6a  $\{X_3,Y_3\}$
\vspace{.5cm}

Orbit coordinates: $\xi=-\arctan(\frac{x^2}{x^1}) -
\arctan(\frac{x^4}{x^3})$, $\chi=-\arctan(\frac{x^2}{x^1}) +
\arctan(\frac{x^4}{x^3})$, $\bar\lambda$

Invariant coordinates: $r=\sqrt{(x^1)^2+(x^2)^2}, R=\sqrt{(x^3)^2+(x^4)^2},
\zeta= e^{i\chi}\lambda$

Invariant $A_\mu$ and $\Psi$:
$$A_1= u_1 \cos(\frac{\xi+\chi}{2})+  u_2\sin(\frac{\xi+\chi}{2}),\
A_2=- u_1 \sin(\frac{\xi+\chi}{2})+  u_2\cos(\frac{\xi+\chi}{2}),$$
$$A_3= u_3 \cos(\frac{\xi-\chi}{2})+  u_4\sin(\frac{\xi-\chi}{2}),\
A_4=- u_3 \sin(\frac{\xi-\chi}{2})+  u_4 \cos(\frac{\xi-\chi}{2}),
\eqno(5.19)
$$
where $u_\mu=u_\mu(r,R)$ and  $\Psi=\psi(r, R, \zeta)$.

 The reduced Lax pair and SDYM equations are deduced from the same
equations obtained in case 4 with the values $\alpha = 1$
and $\beta = 0$ by setting $\p_\chi \psi=0$ and $\p_\chi u_\mu=0$.

\vspace{.5cm}
{\bf 8.} 8a  $\{X_1,X_2,X_3\}$
\vspace{.5cm}

Orbit coordinates: $\phi_1, \phi_2, \phi_3$ in $x= e^{(\phi_1-\phi_2) X_2}
e^{\phi_3 X_1} e^{(\phi_1+\phi_2)X_2} [0,0,0,R]^T$, and $\bar\lambda$

Invariant coordinates: $R=\sqrt{x^\mu x_\mu},  \lambda$

Invariant $A_\mu$ and $\Psi$:
$$[A_1, A_2, A_3, A_4]^T = e^{(\phi_1-\phi_2) X_2} e^{\phi_3 X_1}
e^{(\phi_1+\phi_2) X_2} [u_1, u_2, u_3, u_4]^T, \eqno(5.20a)$$ where
$u_\mu=u_\mu(R).$ For the purpose of the calculations, we can express it in
terms of Cartesian coordinates. We then have:
$$A_\mu = 2({\bar\eta}^a_{\mu\nu} x^\nu v_a + \delta_{\mu\nu} x^\nu v_4),
\eqno(5.20b)$$ where $v_\mu=v_\mu ({\cal R})$, with ${\cal R}=R^2$ and
$v_a=-\frac{1}{2R}u_a\, (a=1,2,3), v_4=\frac{1}{2R} u_4$
 and
$\Psi=\psi({\cal R}, \lambda)$.

Reduced Lax pair:
$${\cal R}^2[\p_{\cal R} + v_4-iv_3 -\lambda(iv_1+v_2)]\psi=0$$
$${\cal R}^2[\lambda\p_{\cal R} - i v_1 + v_2 +
\lambda(iv_3+v_4)]\psi = 0 \eqno(5.21)$$

Reduced SDYM equations:
$$\p_{\cal R} v_1 + \frac{2}{\cal R} v_1 + [v_2,v_3] + [v_1,v_4] = 0$$
$$\p_{\cal R} v_2 + \frac{2}{\cal R} v_2 + [v_1,v_3] - [v_2,v_4] = 0
\eqno(5.22)$$
$$\p_{\cal R} v_3 + \frac{2}{\cal R} v_3 - [v_1,v_2] - [v_3,v_4] = 0$$

Let us add that the Nahm's equations can be retrieved by putting $v_4=0$
and by carrying out the following change of variables:
$\varphi=\frac{-1}{2{\cal R}}$ and $w_a=-2{\cal R}^2 v_a$.
\vspace{.5cm}

{\bf 9.} 9a  $\{Y_1,Y_2,Y_3\}$
\vspace{.5cm}

Orbit coordinates: $\phi_1, \phi_2, \phi_3$ in $x= e^{(\phi_1-\phi_2) Y_2}
e^{\phi_3 Y_1} e^{(\phi_1+\phi_2)Y_2} [0,0,0,R]^T$, and $\bar\lambda$

Invariant coordinates: $R=\sqrt{x^\mu x_\mu}, \zeta = \frac{z^1-\lambda
\bar z^2}{z^2+\lambda \bar z^1}$, where $z^1:=x^1 + i x^2$ and $z^2:=
x^3-ix^4$

Invariant $A_\mu$ and $\Psi$:
$$[A_1, A_2, A_3, A_4]^T = e^{(\phi_1-\phi_2) Y_2} e^{\phi_3 Y_1}
e^{(\phi_1+\phi_2) Y_2} [u_1, u_2, u_3, u_4]^T,
\eqno(5.23a)
$$
where $u_\mu=u_\mu(R)$. In order to facilitate calculations, it can be
rewritten in terms of Cartesian coordinates:
$$A_\mu = 2(\eta^a_{\mu\nu} x^\nu v_a + \delta_{\mu\nu} x^\nu v_4),
\eqno(5.23b)$$ where $v_\mu=v_\mu ({\cal R})$, with ${\cal R}=R^2$ and
$v_\mu=-\frac{1}{2R}u_\mu$ and $\Psi=\psi(R, \zeta)$.

Reduced Lax pair:
$$[\p_{\cal R} +iv_3+v_4 +\zeta(v_2+iv_1)]\psi=0$$
$$[\zeta\p_{\cal R} + i v_1 - v_2 +
\zeta(v_4-iv_3)]\psi = 0 \eqno(5.24)$$

Reduced SDYM equations:
$$[\p_{\cal R} v_1 +[v_2,v_3]-[v_1,v_4]]\psi = 0$$
$$[\p_{\cal R} v_2 +[v_3,v_1]-[v_2,v_4]]\psi = 0 \eqno(5.25)$$
$$[\p_{\cal R} v_3 +[v_1,v_2]-[v_3,v_4]]\psi = 0$$

The Nahm's equations are derived if we require $v_4=0$ and change $v_a$ to
$-v_a$. Contrary to the previous cases, we would like to point out that
even if the lift of the elements of the symmetry algebra is nontrivial, the
reduced Lax pair does not involve vector components in the direction of the
new spectral parameter.

\vspace{.5cm}

{\bf 10.} 13a $\{X_3+Y_3+cP_4\}$
\vspace{.5cm}

Orbit coordinates: $\xi=-\arctan(\frac{x^2}{x^1}) - c x^4, \bar\lambda$.

Invariant coordinates: $r=\sqrt{(x^1)^2+(x^2)^2}, x^3, \chi=-
c\arctan(\frac{x^2}{x^1})+x^4,$ $\zeta = e^{i\gamma \xi} \lambda$, where
$\gamma=\frac{1}{1+c^2}$.

Invariant $A_\mu$ and $\Psi$:
$$A_1= u_1\cos(\gamma \xi)+ u_2\sin(\gamma \xi),\
A_2=- u_1\sin(\gamma \xi)+ u_2\cos(\gamma \xi),\
A_3=u_3,\ A_4=u_4,
\eqno(5.26)
$$
where $u_\mu=u_\mu(r,x^3,\chi)$ and $\Psi=\psi(r, x^3, \chi, \zeta)$.

Reduced Lax pair:
$$[\p_r - i(e^{i\gamma c \chi}\zeta+\frac{c}{r})\p_\chi -
\zeta e^{i\gamma c\chi}\p_3 + (\frac{\gamma}{r}-\gamma c e^{i\gamma c \chi}
\zeta)\zeta\p_\zeta + e^{i\gamma c \chi}((u_1+iu_2) - \zeta(u_3+iu_4))]
\psi = 0$$
$$[e^{i\gamma c \chi} \zeta \p_r + \p_3 + i(e^{i\gamma c \chi} \frac{c}{r}
\zeta - 1) \p_\chi - (\frac{\gamma}{r} \zeta e^{i\gamma c \chi} +\gamma c)
\zeta \p_\zeta +\zeta(u_1-iu_2) + u_3-iu_4]\psi = 0 \eqno(5.27)$$

The above linear system is composed of holonomic vectors if the first
equation is multiplied by $r$.

Reduced SDYM equations:
$$\sin(\gamma c \chi) (\p_ru_1 + \frac{\gamma}{r} u_1 -\frac{c}{r}
\p_\chi u_2) + \cos(\gamma c \chi) (\p_ru_2 + \frac{\gamma}{r} u_2 +
\frac{c}{r} \p_\chi u_1) $$
$$- \p_3 u_4 + \p_\chi u_3 + [u_1,u_2] - [u_3,u_4] = 0$$
$$\cos(\gamma c \chi) (\p_r u_3 - \frac{c}{r} \p_\chi u_4) - \sin(\gamma c
\chi) (\p_r u_4 + \frac{c}{r} \p_\chi u_3) $$
$$- \p_3 u_1 - c \gamma u_1 - \p_\chi u_2 + [u_1,u_3] + [u_2,u_4] = 0
\eqno(5.28)$$
$$\sin(\gamma c \chi) (\p_r u_3 - \frac{c}{r} \p_\chi u_4) + \cos(\gamma c
\chi) (\p_r u_4 + \frac{c}{r} \p_\chi u_3) $$
$$+ \p_3 u_2 + c \gamma u_2 - \p_\chi u_1 - [u_2,u_3] + [u_1,u_4] = 0$$
\vspace{.5cm}

{\bf 11.} 13b $\{X_3+Y_3+cP_4, P_3\}$
\vspace{.5cm}

The reduced Lax pair and SDYM equations are obtained by ignoring any
dependence with respect to the orbit variable $x^3$ in the equations (5.27)
and (5.28).
\vspace{.5cm}

{\bf 12.} 13c $\{X_3+Y_3+cP_4, P_1, P_2\}$
\vspace{.5cm}

Orbit coordinates: $x^1, x^2, \theta=\frac{x^4}{c}, \bar \lambda$

Invariant coordinates: $x^3, \zeta = e^{-i\frac{x^4}{c}} \lambda$

Invariant $A_\mu$ and $\Psi$:
$$A_1=u_1,\
A_2=u_2,\
A_3 = u_3,\ A_4 = u_4,
\eqno(5.29)
$$
where $u_\mu = u_\mu(x^3)$ and $\Psi=\psi(x^3, \zeta)$.

Reduced Lax pair:
$$[\zeta \p_3 + \frac{\zeta^2}{c} \p_\zeta + \zeta (u_3+iu_4) - u_1
-iu_2]\psi = 0$$
$$[\p_3 - \frac{\zeta}{c} \p_\zeta + \zeta (u_1-iu_2) + u_3 - iu_4]\psi = 0
\eqno(5.30)$$

We have a set of holonomic vector fields for this linear system if the
factor $\frac{1}{\zeta}$ is added to the first equation.

Reduced SDYM equations:
$$\p_3 u_1 + \frac{u_1}{c} - [u_1,u_3] - [u_2,u_4] = 0$$
$$\p_3 u_2 + \frac{u_2}{c} - [u_2,u_3] + [u_1,u_4] = 0 \eqno(5.31)$$
$$\p_3 u_4 - [u_1,u_2] + [u_3,u_4] = 0$$

\vspace{1cm}

{\bf VI. CONCLUSION}

\vspace{0.3cm}

In this paper, we have applied the method of reduction by symmetry to
the Lax pair, or linear system, of the SDYM equations on four-dimensional
Euclidean space. Two main aspects to be considered were, first, the
extension of the Lax pair to the product of the Euclidean space and the
space of the spectral parameter ($CP^1$), and second, the lift of the group
action of $SO(5,1)$ to $R^4\times CP^1\subset CP^{3\ast }$ preserving
the complex structures induced on $E^4$ by the linear system (2.4). Using a
classification of the subalgebras of $e(4)$ under conjugacy classes with
respect to the adjoint action of $E_o(4)$, we have reduced the Lax pair for
the SDYM equations under each class representative which produces a
nontrivial residual differential system. A list
of these representatives can be read in Table 1 and the reduced Lax
pairs are given in section V. The compatibility of the reduced Lax pairs
agrees exactly with the similarly reduced SDYM equations. For many
reduced linear systems, a vector component along the (new) spectral
parameter arose, typically when a nontrivial lift of the group action
was involved.

As possible developments of this work, further reductions of the Lax
pairs and SDYM equations can be effected for the representatives of
conjugacy classes of subgroups of $SO(5,1)$ as well as for Yang-Mills
fields ($A_\mu$) invariant up to gauge transformations. One can also carry
out reductions of the same set of equations on $R^{(2,2)}$ with
respect to subgroups of the corresponding conformal group $SO(3,3)$,
and equally for higher-dimensional and self-dual spaces versions of these
equations under subgroups of their space transformation groups. Finally, it
could be interesting to apply the method of symmetry reduction to the
(universal) hierarchy of SDYM equations and to supersymmetric
generalizations of the above systems.

\vspace{1cm}

{\bf  ACKNOWLEDGEMENTS}

\vspace{0.3cm}

Part of this work was done in collaboration with A.D. Popov, Bogoliubov
Laboratory of Theoretical Physics, JINR, Dubna, Moscow Region, Russian
Federation. The author thanks for its kind hospitality the Centre de
Recherches Math\'ematiques of the Universit\'e de Montr\'eal where this
work was completed. This work was supported by  a grant from the National
Sciences and Engineering Research Concil (NSERC) of Canada.

\vspace{1cm}

{\bf REFERENCES}

\begin{enumerate}

\item P.J. Olver, {\it Applications of Lie Groups to Differential
Equations}  (Springer-Verlag, New York, 1986).

\item G.W. Bluman and S. Kumei, {\it Symmetries and Differential Equations}
(Springer-Verlag, New York, 1989).

\item P. Winternitz, in {\it Partially Integrable Evolution Equations in
Physics}, Eds.: R. Conte and N. Boccara, NATO ASI series C, Vol. 310
(Kluwer  Academic Publ., 1990), p. 515.

\item P. Forg\'acs and N.S. Manton, Commun. Math. Phys. {\bf 72}, 15
(1980).

\item J. Harnad, S. Shnider and L. Vinet, J. Math. Phys. {\bf 21}, 2719
(1980).

\item J. Harnad, S. Shnider and J. Tafel, Lett. Math. Phys. {\bf 4}, 107
(1980).

\item R. Jackiw and N.S. Manton, Ann. Phys. {\bf 127}, 257 (1980).

\item V. Hussin, J. Negro and M.A. del Olmo, Ann. Phys. {\bf 231}, 211
(1994).

\item L. Vinet, Phys. Rev. {\bf D24}, 3179 (1981).

\item M.F. Atiyah, {\it Classical Geometry of Yang-Mills Fields} (Scuola
 Normale Superiore, Pisa, 1979).

\item R.S. Ward and R.O. Wells Jr, {\it Twistor Geometry and Field Theory}
(Cambridge University Press, Cambridge, 1990).

\item R.S. Ward, {\it Twistors in Mathematics and Physics}, Eds.: T.N.
Bailey and R.J. Baston, London Mathematical Society Lecture Note Series
156, p. 246 (Cambridge University Press, Cambridge, 1990).

\item M.J. Ablowitz and P.A. Clarkson, {\it Solitons, Nonlinear Evolution
Equations and Inverse Scattering} (Cambrige University Press,
Cambridge, 1991).

\item R.S. Ward, Phys. Lett. {\bf A112}, 3 (1985).

\item T.A. Ivanova and A.D. Popov, Lett. Math. Phys. {\bf 23}, 29 (1991).

\item S. Chakravarty, S. Kent and E.T. Newman, J. Math. Phys. {\bf 33}, 382
(1992).

\item I. Bakas and D.A. Depireux, Mod. Phys. Lett. {\bf A6}, 399 (1991).

\item I. Bakas and D.A. Depireux, Mod. Phys. Lett. {\bf A6}, 1561 (1991).

\item T.A. Ivanova and A.D. Popov, Phys. Lett. {\bf A170}, 293 (1992).

\item L.J. Mason and G.A.J. Sparling, Phys.Lett. {\bf A137}, 29 (1989).

\item L.J. Mason and G.A.J. Sparling, J.Geom.Phys. {\bf 8}, 243 (1992).

\item S. Chakravarty and M.J. Ablowitz, {\it Painlev\'e Transcendents,
their Asymptotics and Physical Applications}, Eds.: D. Levi and P.
Winternitz, NATO ASI Series B, Vol. 278, p. 331 (Plenum Press, New York,
1992).

\item M.J. Ablowitz, S. Chakravarty and L.A. Takhtajan, Commun. Math. Phys.
{\bf 158}, 289 (1993).

\item R.S. Ward, Nucl. Phys. {\bf B236}, 381 (1984).

\item T.A. Ivanova and A.D. Popov, Lett. Math. Phys. {\bf 24}, 85 (1992).

\item I.A.B. Strachan, J. Math. Phys. {\bf 34}, 293 (1993).

\item J. Tafel, J. Math. Phys. {\bf 34}, 1892 (1993).

\item A.A. Belavin and V.E. Zakharov, Phys. Lett. {\bf B73}, 53 (1978).

\item R.S. Ward, Phys. Lett. {\bf A61}, 81 (1977).

\item J. Fletcher and N.M.J. Woodhouse, {\it Twistors in Mathematics and
Physics}, Eds.: T.N. Bailey and R.J. Baston, London Mathematical Society
Lecture Note Series 156, p. 260 (Cambridge University Press, Cambridge,
1990).

\item N.M.J. Woodhouse and L.J. Mason, Nonlinearity {\bf 1}, 73 (1988).

\item L.J. Mason and N.M.J. Woodhouse,  Nonlinearity {\bf 6}, 569 (1993).

\item M. Legar\'e and A.D. Popov, Phys. Lett. {\bf A198}, 195 (1995).

\item T.A. Ivanova and A.D. Popov, Pis'ma Zh. Eksp. Teor. Fiz. {\bf 61},
142 (1995) (JETP Lett. {\bf 61}, 150 (1995)).

\item T.A. Ivanova and A.D. Popov, Teor. Mat. Fiz. {\bf 102}, 384 (1995).

\item A. Mendoza, A. Restuccia, and I. Martin, Lett. Math. Phys. {\bf 21},
221 (1991).

\item S.J. Gates Jr. and H. Nishino, Phys. Lett. {\bf B299}, 255 (1993).

\item H. Ooguri and C. Vafa, Nucl. Phys. {\bf B367}, 83 (1991), and
references therein.

\item H. Nishino and S.J. Gates Jr., Mod. Phys. Lett. {\bf A7}, 2543
(1992).

\item N.M.J.Woodhouse, Class. Quantum Grav. {\bf 2}, 257 (1985).

\item M.K. Prasad, Physica {\bf D1}, 167 (1980).

\item A.Lichnerowicz, {\it G\'eom\'etrie des Groupes de Transformations}
 (Dunod, Paris, 1958).

\item S.Kobayashi and K. Nomizu, {\it Foundations of Differential Geometry,
Vol. II} (John Wiley \& Sons, New York, 1969).

\item M. Kovalyov, M. Legar\'e and L. Gagnon, J. Math. Phys. {\bf 34}, 3425
(1993).

\item J. Patera, P. Winternitz and H. Zassenhaus, J. Math. Phys. {\bf 16},
1597 (1975).

\item L. Gagnon, Can. J. Phys. {\bf 67}, 1 (1989).

\item V. Hussin, P. Winternitz and H. Zassenhaus, Linear Algebra Appl. {\bf
141}, 183 (1990).

\item E.B. Bogomolny, Sov. Nucl. Phys. {\bf 24}, 861 (1976).

\item M.K. Prasad and C.M. Sommerfield, Phys. Rev. Lett. {\bf 35}, 760
(1975).

\item R.S. Ward, Phil. Trans. Soc. Lond. {\bf A315}, 451 (1985).

\item W. Nahm, in {\it Monopoles in Quantum Field Theory} eds : N.S.
Craigie, P. Goddard and W. Nahm (World Scientific, Singapore, 1981) p. 87.

\item N.J. Hitchin, Commun. Math. Phys. {\bf 120}, 613 (1983).

\item N.J. Hitchin, {\it Monopoles, Minimal Surfaces and Algebraic Curves},
S\'eminaire de Math\'ematiques Sup\'erieures 105 (Les Presses de
l'Universit\'e de Montr\'eal, Montr\'eal, 1987).

\end{enumerate}

\newpage
\centerline {\bf TABLE 1.}
\centerline{\bf Representatives of Conjugacy Classes of Subalgebras of
$e(4)$}
\centerline{\bf Leading to Nonlinear Reduced SDYM Equations$^{\dag}$}
\vspace{.5cm}
\halign{\indent\hfil#\hfil&\hskip 1.5cm#\hfil&\hskip
1.5cm\hfil#\hfil\cr {\bf Representative(s) \# in ref. 44}&{\bf Basis of
Subalgebra}&{\bf Condition}\cr
  &  &  \cr
1a&$P_4$& \cr 1b&$P_3, P_4$& \cr
  &  &  \cr
1c&$P_1, P_2, P_3$& \cr
  &  &  \cr
2a,3a,4a,5a&$\alpha X_3 + \beta Y_3$&$\alpha,\beta \in
R$\cr
  &  &  \cr
2b,3b,4c,5b&$\alpha X_3 + \beta Y_3,P_3,P_4$&$\alpha,\beta \in R$\cr
  &  &  \cr
4b&$X_3+Y_3, P_3$&\cr
  &  &  \cr
6a&$X_3, Y_3$&\cr
  &  &  \cr
8a&$X_1, X_2, X_3$&\cr
  &  &  \cr
9a&$Y_1, Y_2, Y_3$&\cr
  &  &  \cr
13a&$X_3+Y_3+c P_4$&$c\in R$\cr
  &  &  \cr
13b&$X_3+Y_3+c P_4, P_3$&$c\in R$\cr
  &  &  \cr
13c&$X_3+Y_3+c P_4,P_1, P_2$&$c\in R$\cr}

\vspace{1cm}

{\dag}{The reduced SDYM equations are not equivalent to a zero
curvature condition without parameter, or to a system of uncoupled linear
first order O.D.E.'s.}

\end{document}